# Role of coexisting ferromagnetic and antiferromagnetic phases on the magnetocaloric effect in metamagnetic $Tb_2Ni_2Sn$

Pramod Kumar, Niraj K. Singh, K. G. Suresh[a]

*Department of Physics, I. I. T .Bombay, Mumbai 400076, India*

A. K. Nigam

*Tata Institute of Fundamental Research, Homi Bhabha Road, Mumbai 400005, India*

Abstract

We report the anomalous magnetocaloric behavior in the polycrystalline compound $Tb_2Ni_2Sn$. The magnetization measurements show that this compound shows multiple magnetic transitions, which are attributed to the coexistence of ferromagnetic and antiferromagnetic phases at low temperatures. With increase in field and temperature, the compound undergoes a metamagnetic transition to a ferromagnetic state. In the temperature range where the antiferromagnetic phase is dominant, it exhibits inverse (negative) magnetocaloric effect. At temperatures close to the Neel temperature, the compound shows positive magnetocaloric effect. Below the critical field needed for the metamagnetic transition, the temperature variation of the magnetocaloric effect is seen to be correlated with the ferromagnetic fraction.

Currently, there is a great deal of interest in utilizing the magnetocaloric effect (MCE) as an alternative technology for refrigeration from the ambient temperature to the sub-Kelvin range of temperatures, replacing the common gas compression/expansion technology. Magnetic refrigeration in the near-room temperature (~300 K) regime is of particular interest because of the potential impact on energy efficiency and environmental concerns. Materials with giant MCE are needed to improve the efficiency of magnetic refrigerators. Magnetic refrigeration is based on the magnetocaloric effect, which is the ability of some magnetic materials to heat up when they are magnetized, and cool down when removed from the magnetic field[1]. The thermodynamic quantities that characterize MCE are the isothermal magnetic entropy change ($\Delta S_M$) and the adiabatic temperature change ($\Delta T_{ad}$)[2]. Therefore improvements in magnetic refrigeration technology require detailed investigations on the microscopic mechanisms that are responsible for enhancing these two quantities. Since Brown[3] described a practical near-room temperature magnetic refrigerator using magnetocaloric effect, and more recently, the discovery[4] of the giant magnetocaloric effect in $Gd_5(Si_2Ge_2)$, the interest in this research area has been considerably enhanced[5,6].

The variety of magnetic phenomena exhibited by many rare-earth (R)–transition-metal (TM) intermetallic compounds render them suitable for applications, specially those based on MCE. Among the R-TM intermetallics, $R_2Ni_2Sn$ compounds remain unexplored with respect to their magnetocaloric behavior. These compounds show interesting multiple magnetic transitions which may render them as good magnetic refrigerants. These compounds, in general, crystallize in the orthorhombic $W_2CoB_2$-type structure. We have taken up a detailed study on the magnetic and magnetocaloric properties of these compounds with various rare earths. In this paper, we report the results on the magnetocaloric effect in polycrystalline $Tb_2Ni_2Sn$ compound. The magnetic structure of $Tb_2Ni_2Sn$ is determined mainly by (i) the nature of the RKKY interactions and (ii) crystalline electric fields (CEF) which determines the ground state of Tb ion[7]. The oscillatory character of the RKKY interaction leads to the occurrence of commensurate or incommensurate magnetic structures. CEF effects may cause anisotropic exchange interactions, which may also play a role in determining the magnetic state at various temperatures.

Polycrystalline sample of $Tb_2Ni_2Sn$ was synthesized by conventional method of arc melting. The as-cast sample was annealed under vacuum at 800 °C for 3 weeks. The phase purity was checked by the Rietveld refinement of the powder x-ray diffraction (XRD) data and SEM-EDAX analysis. The magnetization (M) was measured using a VSM (Oxford instruments) in the temperature (T) range of 2-240K, up to a field (H) of 100 kOe. Thermo magnetic analysis was performed both under zero-field cooled (ZFC) and field-cooled (FC) modes. In the former case, the sample was cooled in the absence of a field and the magnetization was measured during warming, by applying a nominal field of 500 Oe. In the FC mode, the sample was cooled in presence of a field and the magnetization was measured during warming, under the four different fields of 500Oe, 20kOe, 40kOe and 60kOe. The magnetocaloric effect has been calculated in terms of isothermal magnetic entropy change, using the magnetization isotherms collected over a wide range of temperatures.

Fig.1 shows the Rietveld fitted XRD plot, which shows that the compound has formed in single phase with the $W_2CoB_2$-type orthorhombic structure (space group=*Immm*). In this context, it is of relevance to mention that Chevalier et al.[8] have reported that $R_2Ni_2Sn$ compounds with R=Tb and Dy are always accompanied by impurity phases namely RNiSn and $RNi_2$. In order to investigate the presence of such phases in our sample, we have carried out composition analysis using SEM-EDAX. The composition analysis has shown that the sample is single phase with the 2:2:1 (Tb:Ni:Sn) composition throughout. In the light of these results, we rule of any impurity phases in our sample. The lattice parameters obtained from the Rietveld refinement are $a$=4.2770(3) Å, $b$=5.6085(4) Å and $c$=8.3228(5)Å. We find that the value of $a$ is very close to that obtained by Chevalier et al.[8], but $b$ and $c$ values show some difference as compared to the reported values.

Fig. 2 shows the M-T plot of $Tb_2Ni_2Sn$ collected in an applied field of 500 Oe, under ZFC and FC conditions. The transition temperature is calculated from the d$M$/d$T$ vs T plot, which shows that this compound exhibits three magnetic transitions corresponding to the maxima at about $T_1 = 8$ K, $T_2 = 42$ K and $T_N = 66$ K. The susceptibility exhibits Curie-Weiss behavior above 80 K with a paramagnetic Curie temperature ($\theta_P$) = 39 K and effective magnetic moment ($\mu_{eff}$) =7.7 $\mu_B$/Tb. It is to be noted that the values of $T_1$, $T_2$, $T_N$ and $\theta_P$ obtained in the present case are in good agreement with the values reported by Chevalier et al. However, the value of $\mu_{eff}$ obtained by us is much less compared to the value of 10.2 $\mu_B$/Tb obtained by Chevalier et al. In our case, the paramagnetic moment is about 21% smaller than the free ion value of $Tb^{3+}$, which may be due to the crystal field effects. Inset of fig. 2 shows the temperature dependence of the FC magnetization data in applied fields of 20 kOe, 40 kOe and 60 kOe. It can be seen that in fields higher than 500 Oe, only one magnetic transition, i.e. at $T_N$, could be observed. In addition, this transition shifts towards lower temperature with increase in the field. This type behavior confirms the antiferro-para magnetic nature transition at 66 K. Penc et al.[9] have observed two magnetic transitions in the temperature dependence of ac susceptibility of $Nd_2Ni_2Sn$ and $Dy_2Ni_2Sn$.

Fig.3(a) shows the variation of magnetization for both increasing and decreasing magnetic fields at 5 K. As is evident from this figure, the low field (~20 kOe) part of the M-H plot is quite linear suggesting the presence of antiferromagnetic (AFM) phase. With further increase in field, the compound undergoes a metamagnetic transition from the AFM state to ferromagnetic (FM) state, resulting in high magnetization. The nature of the plot suggests that for $H<H_{cr1}$[Region (a)] the compound shows the antiferromagnetic (AFM) nature. In the range of $H_{cr1}<H<H_{cr2}$ [Region (b)] compound shows the AF+AFM type mixed behavior. When applied field $H>H_{cr2}$ [Region (c)], the system tends to become completely ferromagnetic. It can be seen that in the region (c) the magnetization is nearly saturated at the highest applied field and the remanence is almost negligible, which indicates that the compound is magnetically soft. The saturation magnetic moment obtained from the M vs. H plot at 5 K is found to be 8.2$\mu_B$ per $Tb^{3+}$ ion whereas the theoretical $g_J J$ value of $Tb^{3+}$ ion is 9 $\mu_B$. The lower value in the present case may also be due to the crystal field effects. In the case of other $R_2Ni_2Sn$ compounds also, a similar behavior has been reported[9].

Figure 3(b) shows the magnetization as a function of applied field at different temperatures. These isotherms have been obtained by warming up the samples to temperatures well above $T_N$, after recording each isotherm. It is of interest to note that in the low field (H< $H_{cr1}$) region, the magnetization increases with temperature whereas the trend reverses for the high field (H>$H_{cr2}$) region. This effect continues up to critical temperature ($T_{cr}$ ~ 55K). On the basis of these factors, it is clear that the percentage of the ferromagnetic component increases gradually with increase in temperature. The values of critical fields $H_{cr1}$ and $H_{cr2}$ decrease with temperature as shown in the inset of fig.3(b).

The magnetocaloric effect in terms of the magnetic entropy change, $\Delta S_M(T, \Delta H)$, has been estimated from the magnetization isotherms, which were measured with a temperature interval of 5 K, for a maximum field of 80 kOe. The $\Delta S_M(T, \Delta H)$ values were calculated by numerically integrating the Maxwell equation,

$$\Delta S_m(T_{av,i}, H_2) = \frac{1}{T_{i+1} - T_i} \int_0^{H_2} (M(T_{i+1}, H) - M(T_i, H)) dH$$

where $T_{av,i}$ is the average of $T_i$ and $T_{i+1}$[10,11,12,13].

Fig.4 shows the variation of isothermal magnetic entropy change (-$\Delta S_M$) for Tb$_2$Ni$_2$Sn as a function of temperature, for several values of H (=$\Delta H$) namely, 20, 40, 60 and 80 kOe. In contrast to the usual observation of a single MCE peak, in this case, the temperature dependence of MCE is quite different. The entropy change is positive (negative MCE) below ~60 K and it changes its sign and becomes negative (positive MCE) at higher temperatures, giving rise to a distinguishable minimum and a maximum for all fields above 20 kOe. It is to be noted that while the temperatures corresponding to the minimum and zero entropy change shift to lower temperatures with increase in field, the position of the maximum is almost insensitive to the field. This type of behavior can be analyzed in terms of the ferromagnetic(FM)-antiferromagnetic(AFM) phase coexistence and the variation in the ratio of these phases under different applied fields. Table 1 shows the maximum and minimum values of ($-\Delta S_M$) with the corresponding temperatures and fields.

Table.1 Maxima and minima values of $-\Delta S_M$ along with the corresponding temperatures ($T^{max}$ and $T^{min}$) and fields.

| H (kOe) | $T^{max}$ (K) | $-\Delta S_M^{max}$ (J/kg K) | $T^{min}$ (K) | $-\Delta S_M^{min}$ (J/kg K) |
|---|---|---|---|---|
| 20 | 68.9 | 0.9 | 57.5 | -1.1 |
| 40 | 68.7 | 2.8 | 52.2 | -4.5 |
| 60 | 67.6 | 5.8 | 39.5 | -8.1 |
| 80 | 67.3 | 9.1 | 37.2 | -7.5 |

Inset of Fig. 4 shows the variation of the entropy change with applied field for several temperatures in the range of 12.5 -78.5 K. As is evident from this figure, at the lowest temperature(which is close to $T_1$), MCE is negligibly small and almost independent of the field. In the temperature range of $T_1 - T_2$, the MCE is negative. This observation clearly indicates that the compound is predominantly antiferromagnetic in this region of temperature. When temperature is just above $T_N$, the MCE is positive and increases monotonically with field.

Fig.5. shows the ferromagnetic fraction as a function of temperature for several values of H. The ferromagnetic fraction has been calculated by subtracting the magnetization arising from the antiferromagnetic component (linear part of the magnetization

isotherms) from the saturated magnetization values. The fraction of coexisting phases (AFM+FM) can be controlled by the value of H. In H=20 kOe case, the percentage of AFM phase is considerable because this field is not sufficiently large to reach the onset of the metamagnetic transition. For higher fields, the FM fraction increases initially and then decreases near $T_N$. The inset of fig. 5 shows the temperature variation of first derivative of the ferromagnetic fraction as well as the isothermal magnetic entropy change calculated for 40 kOe. It is quite evident that these two variations are correlated. Furthermore, figures 4 and 5 show that the MCE minima and maxima correspond to the points of inflection of the increasing and decreasing parts of the curve showing ferromagnetic fraction, respectively. This feature is in agreement with the prediction of Tishin et. al[14]. The increasing part of these curves reflects the dominance of the Zeeman energy over the thermal energy, while the decreasing portion of this curves shows the reversed scenario. Further increasing tendency seen in the ferromagnetic fraction at high temperatures is due to the paramagnetic contribution as this temperature range is above $T_N$.

The magnetic and magnetocaloric results obtained in $Tb_2Ni_2Sn$ show some striking similarities with those seen in $(Pr,Ca)MnO_3$ manganate system reported by Gomes et al.[15] recently. Both these systems show the coexistence of antiferromagnetic and ferromagnetic components in the low temperature range. As a consequence, in both these systems, the sign of MCE changes from negative to positive as the temperature is increased. Furthermore, the temperature variation of magnetic entropy change is closely related o the change in the ferromagnetic fraction. It has been suggested that the anomalous MCE behavior in (Pr,Ca)MnO$_3$ is due to the presence of successive ferromagnetic and antiferromagnetic slabs in the unit cell. Von Ranke et al.[16] have shown that the MCE in DyAl$_2$ single crystals show a similar negative to positive change over behavior, which is attributed to the metamagnetic transition. Very recently, a similar MCE behavior has been reported in Ce(Fe,Ru)$_2$ compounds, which is explained on the basis of the coexisting FM and AFM phases[17].

Based on magnetic and neutron diffraction studies in Dy$_2$Ni$_2$Sn which is iso-structural to Tb$_2$Ni$_2$Sn, Penc et al.[9] have shown that the low temperature magnetic structure of this compound is noncollinear with large AFM and small FM components. The FM component is along the *a*-axis while the AFM component is in the *ac*-plane. These studies have also shown that Ni is nonmagnetic. The magnetic transitions seen in these compounds arise due to the anisotropic exchange interactions. The anisotropy arises from the considerably different intra-planar and inter planar R-R bond lengths. Therefore, it is quite possible that the R sublattice in Tb$_2$Ni$_2$Sn resembles that of (Pr,Ca)MnO$_3$ with different types of magnetic order within and between the layers in the unit cell. Our observations suggest that at low temperatures (below T$_1$), Tb$_2$Ni$_2$Sn possesses predominantly AFM phase which decreases with increase in temperature and (or) applied field (as seen in Fig. 5). Therefore, the region T$_1$<T<T$_2$ is the mixed phase region. In the region T$_2$<T<T$_N$, the reaming AFM order completely disappears, resulting in the completely paramagnetic state above T$_N$.

In conclusion, we find that the magnetocaloric effect seen in polycrystalline Tb$_2$Ni$_2$Sn is quite interesting. The field-induced sign change of the magnetic entropy could be exploited for applications such as magnetic refrigerators as well as heat pumps. A detailed neutron diffraction or local Hall probe measurement is essential to probe the exact magnetic structure and the expected co-existence of FM/AFM phases in this compound.

**Acknowledgement**

One of the authors (KGS) thanks I.R.S.O., Govt. of India for proving financial support for this work.

[9] Penc B., Hofmann M., Starowicz M. and Szytula A., 1999 *J. Alloys and Compounds*.**282**, 8.

[10] Singh N. K., Suresh K.G., Nigam A.K., 2003 *Solid State Commun.* **127**, 373 .

[11] Singh N. K., Agarwal S., Suresh K.G., Nirmala R., Nigam A. K. , Malik S. K., 2005 *Phys. Rev. B* **72**, 014452 .

[12] Kumar Pramod, Singh N. K., Suresh K.G. , Nigam A.K., *J. Alloys and Comp.*(in press).

[13] Singh N. K., Kumar Pramod, Suresh K. G., Nigam A. K., Coelho A. A. and Gama S., *Phys. Rev. B.* (communicated).

[14] Tishin A. M., Gschneidner K. A., and Pecharsky V. K., 1999 *Phys. Rev. B* **59**, 503 .

[15] Gomes A. M., Garcia F., Guimarães A. P., Reis M. S., Amaral V. S., 2004 Appl. Phys. Lett. **85**, 4974.

[16] Ranke P. J. von, Oliveira I. G. de, Guimaraes A. P. and Silva X. A. De, 2000 *Phys. Rev. B.* **61**, 447.
[17] Chattopadhyay M.K., Manekar M A and Roy S B  2006 *J. Phys. D: Appl. Phys.* **39**, 1006.
**LIST OF FIGURES**

Fig.1 Rietveld refined powder x-ray diffractograms of $Tb_2Ni_2Sn$. The plot at the bottom shows the difference between the experimental and calculated intensities.

Fig. 2 Temperature variation of ZFC and FC magnetization of $Tb_2Ni_2Sn$ in a field of 500 Oe. The inset shows the variation of FC magnetization in fields of 20, 40 and 60kOe.

Fig.3(a) M-H isotherms of $Tb_2Ni_2Sn$ compound at 5 K

Fig.3(b) M-H isotherms of $Tb_2Ni_2Sn$ in the temperature range of 10-60 K. The inset shows the variation of critical fields with temperature.

Fig.4 Variation of isothermal magnetic entropy of $Tb_2Ni_2Sn$ with temperature, for a field change of 20, 40, 60 and 80 kOe. The inset shows the variation of entropy change with field for different temperatures.

Fig.5 Ferromagnetic fraction as a function of temperature for different applied fields. The inset shows the comparison between $-\Delta S_M$ and derivative of the ferromagnetic fraction, for a field of 40 kOe.

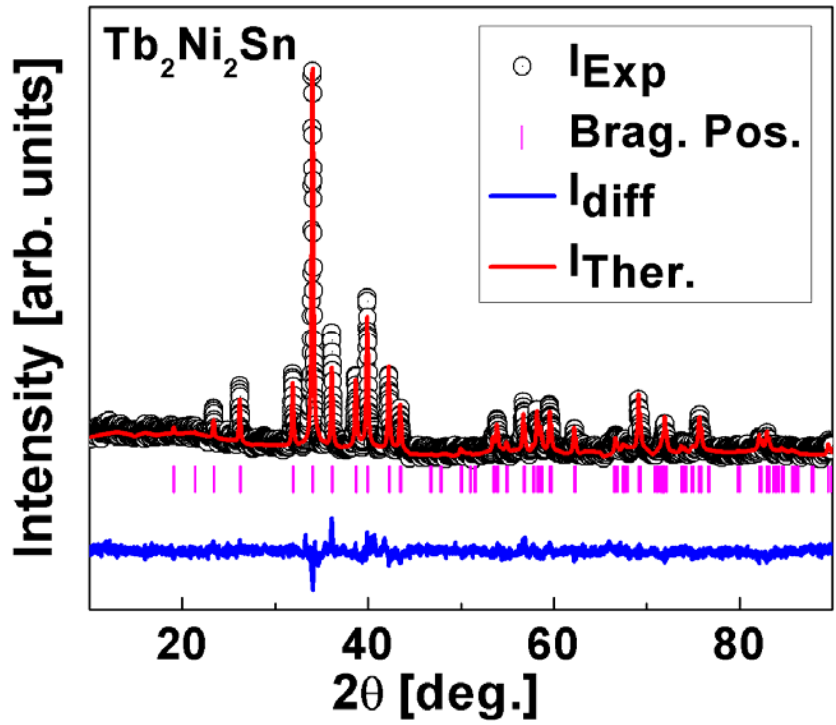

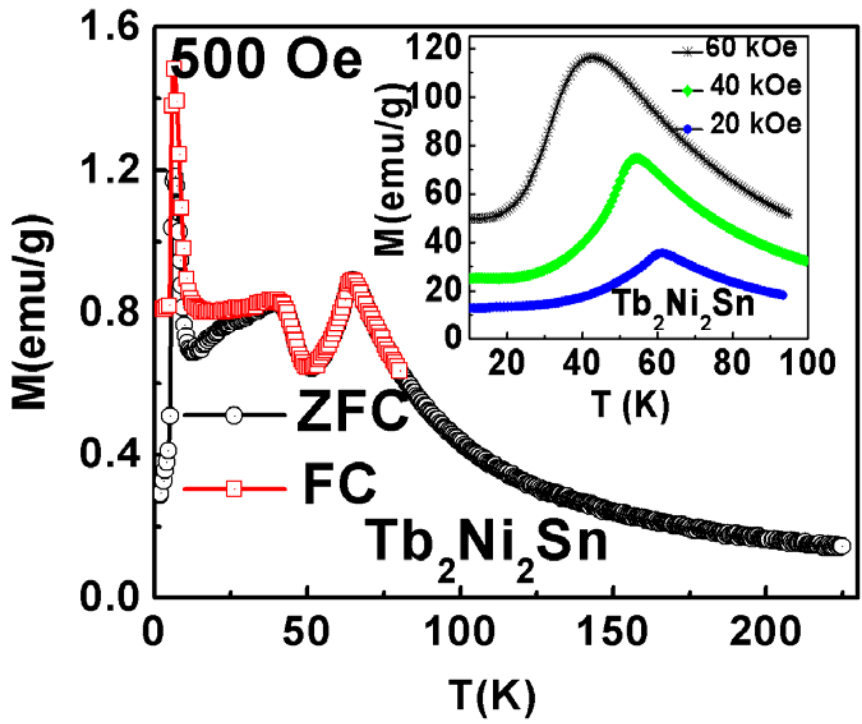

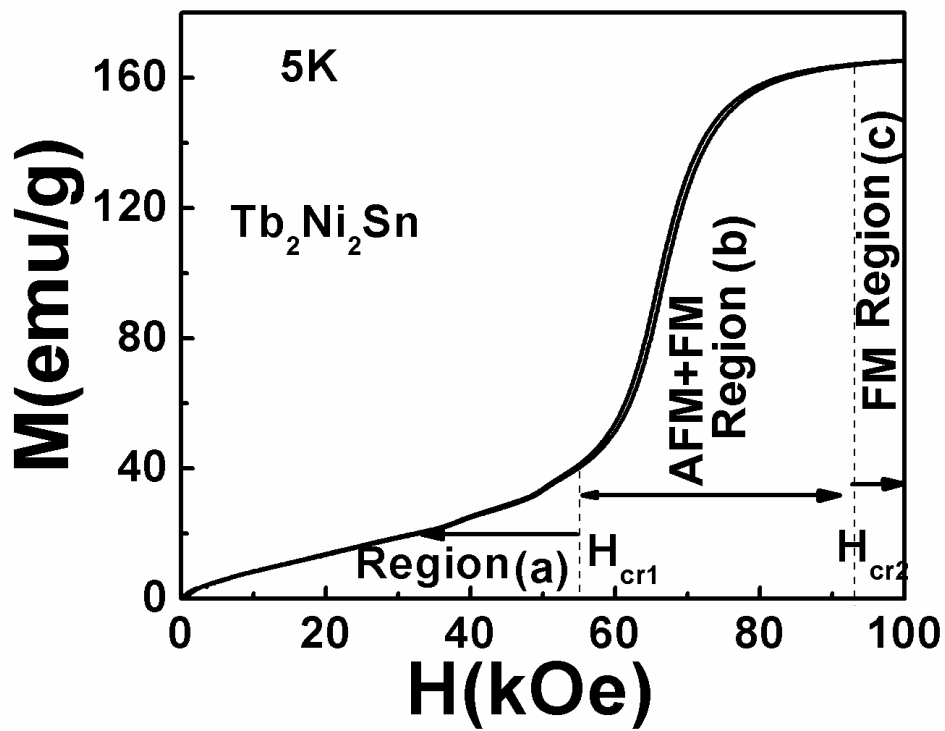

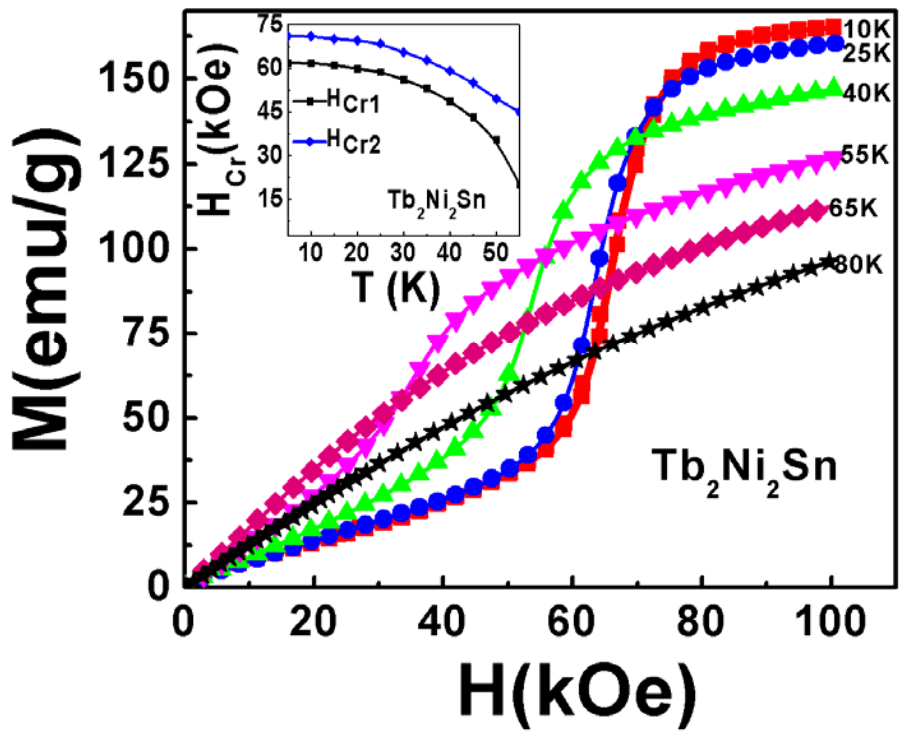

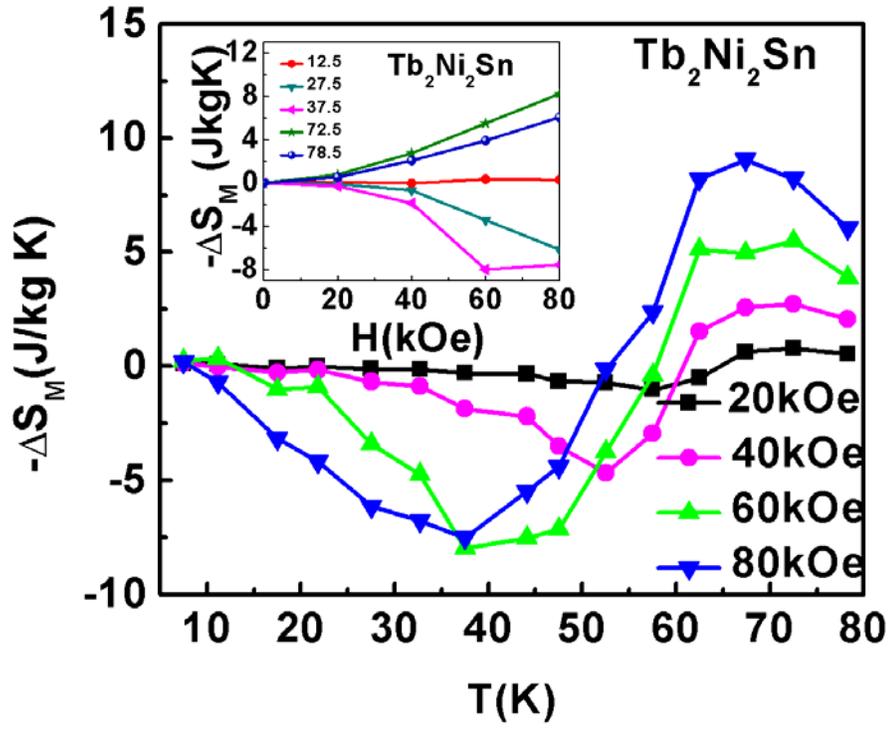

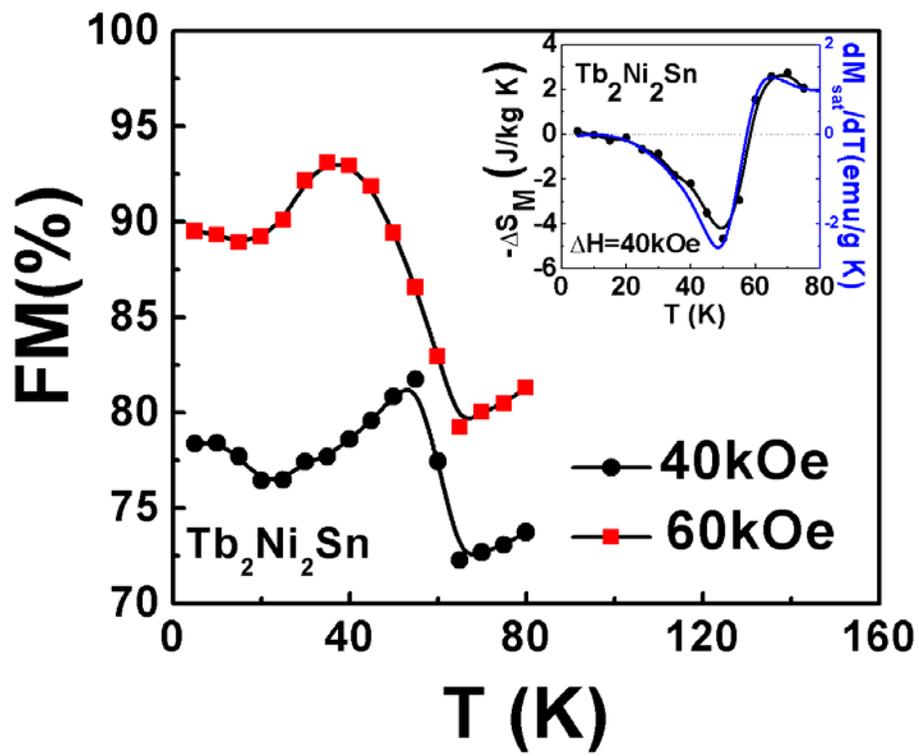